\documentclass[conference]{IEEEtran}
\IEEEoverridecommandlockouts
\usepackage{cite}
\usepackage{amsmath,amssymb,amsfonts}
\usepackage{algorithmic}
\usepackage{bm}
\usepackage[tableposition=top]{caption}
\usepackage{float}
\floatstyle{plaintop}
\restylefloat{table}
\usepackage{svg}
\usepackage{pgf}
\usepackage{graphicx}
\usepackage{textcomp}
\usepackage{xcolor}
\usepackage{hyperref}
\usepackage{import}
\usepackage{todonotes}
\usepackage{color}
\usepackage{multirow}
\usepackage{soul}
\usepackage{subfig}
\usepackage{gensymb}
\usepackage{footnote}
\usepackage{float}
\restylefloat{table}
\makesavenoteenv{tabular}
\usepackage{float}
\usepackage{makecell}
\usepackage{booktabs}
\usepackage{capt-of}
\usepackage[flushleft]{threeparttable}

\def\BibTeX{{\rm B\kern-.05em{\sc i\kern-.025em b}\kern-.08em
    T\kern-.1667em\lower.7ex\hbox{E}\kern-.125emX}}

\begin{document}

\title{MaxEVA: \underline{Max}imizing the \underline{E}fficiency of Matrix Multiplication on \underline{V}ersal \underline{A}I Engine}

\author{
\IEEEauthorblockN{
Endri Taka\IEEEauthorrefmark{1}, 
Aman Arora\IEEEauthorrefmark{1}\IEEEauthorrefmark{2}, 
Kai-Chiang Wu\IEEEauthorrefmark{1}\IEEEauthorrefmark{3}, 
and Diana Marculescu\IEEEauthorrefmark{1}}
\IEEEauthorblockA{
\IEEEauthorrefmark{1}\emph{The University of Texas at Austin, USA}, \IEEEauthorrefmark{2}\emph{Arizona State University, USA}}
\IEEEauthorblockA{
\IEEEauthorrefmark{3}\emph{National Yang Ming Chiao Tung University, Taiwan}}
\{endri.taka, dianam\}@utexas.edu, aman.kbm@asu.edu, kcw@cs.nycu.edu.tw}

\maketitle
\begin{abstract}
The increasing computational and memory requirements of Deep Learning (DL) workloads has led to outstanding
innovations in hardware architectures.
An archetype of such architectures is the novel Versal AI Engine (AIE) by AMD/Xilinx.
The AIE comprises multiple programmable processors optimized for vector-based algorithms.
An AIE array consisting of 400 processor cores, operating at 1.25 GHz is able to deliver a peak throughput of 8 TFLOPs for 32-bit floating-point (fp32), and 128 TOPs for 8-bit integer (int8) precision.
In this work, we propose \textit{MaxEVA}: a novel framework to efficiently map Matrix Multiplication (MatMul) workloads on Versal AIE devices.
Our framework maximizes the performance and energy efficiency of MatMul applications by efficiently exploiting features of the AIE architecture and resolving performance bottlenecks from multiple angles. 
When demonstrating on the VC1902 device of the VCK190 board, MaxEVA accomplishes up to 5.44 TFLOPs and 77.01 TOPs throughput for fp32 and int8 precisions, respectively.
In terms of energy efficiency, MaxEVA attains up to 124.16 GFLOPs/W for fp32, and 1.16 TOPs/W for int8.
Our proposed method substantially outperforms the state-of-the-art approach by exhibiting up to 
2.19$\times$ throughput gain and \textbf{20.4\%} higher energy efficiency.
The MaxEVA framework provides notable insights to fill the knowledge gap in effectively designing MatMul-based DL workloads on the new Versal AIE devices.
\end{abstract}


\section{Introduction}
\label{sec:Introduction}



Contemporary Deep Learning (DL) workloads present exceptionally high compute demands, with a rate of increase of 1.5$\times$ per year \cite{TPUv42021}.
To keep pace with this explosion, several hardware acceleration solutions have been proposed.
These solutions include GPUs \cite{NVIDIA_V100, NVIDIA_A100, NVIDIA_H100}, FPGAs \cite{AutoSA_FPGA21, Sextans_FPGA22, OpenCL_Arria_FPGA17} and ASICs \cite{Eyeriss_ISCA16, ShiDianNao_ISCA15, NVDLA}, while offering orders of magnitude higher performance and energy efficiency compared to general-purpose CPUs \cite{DSA_Bill_Dally_2020, Patterson_golden_age2019, TPU_paper2017}.
Among all solutions, FPGAs are an appealing candidate for DL
because of their reconfigurability.
More recently, to keep up with the demands of DL workloads, FPGA architectures have become more DL-specialized \cite{intel_tensor_block, Versal_AI_Engines_FPGA_paper, tensor_slice}.
To this end, AMD/Xilinx released the Versal Adaptive Compute Acceleration Platform (ACAP), which features the novel AI Engine (AIE) processors.
The Versal ACAP is a heterogeneous system-on-chip (SoC), comprising the AIEs along with the reconfigurable logic (FPGA) and scalar processors (CPUs) \cite{Versal_AI_Engines_FPGA_paper, Alok2020apocalypse}.
The AIE consists of multiple software programmable processors, specifically optimized for 
DL applications \cite{Vesral_AIE_white_paper}. 




The Versal AIE signifies a new era in reconfigurable computing, while achieving considerably higher performance and energy efficiency in DL workloads compared to traditional FPGA designs\cite{Vesral_AIE_white_paper, charm2023fpga}.
However, the complex AIE architecture poses several new design challenges as well.
The efficient design 
and mapping 
of DL applications on AIE is a non-trivial task.
To address these 
design challenges, we propose the novel MaxEVA framework.
MaxEVA constitutes a systematic methodology to maximize the performance and energy efficiency of Matrix Multiplication (MatMul) 
applications on Versal AIEs.
MaxEVA efficiently utilizes attributes of the AIE architecture (local memory sharing, static circuit-switching, broadcasting), and effectively addresses design challenges that lead to sub-optimal performance (limited I/O and switch bandwidth, reduced AIE-FPGA interface tiles, routing congestion).


In this work, we conduct a comprehensive exploration of using the AIE architecture to optimize MatMul-based DL workloads.
We focus on optimizing MatMul operations, because MatMul is the heaviest compute-bound kernel in many DL workloads, occupying up to 90\% of the execution time \cite{fathom_harvard_16}.
All other memory-bound kernels used in DL, \textit{e.g.}, softmax, layernorm, can be effectively overlapped while MatMul kernels are executing, showing minimal contribution to overall throughput and power consumption \cite{charm2023fpga}.
Moreover, we target both 8-bit integer (int8) and  IEEE 32-bit floating-point (fp32) data types, which are the most commonly used in DL \cite{TPUv42021}.
In summary, the main contributions of this work are:
\begin{itemize}
  \item An optimization methodology based on analytical modelling to maximize the performance of MatMul on Versal AIE. Our methodology is generalizable to any Versal AIE device and addresses various performance bottlenecks, leading to maximal utilization of the AIE resources.
  \item A sophisticated AIE kernel placement strategy to effectively leverage the most efficient data movement mechanisms of the Versal AIE architecture.
  \item Demonstration of the MaxEVA framework on the VC1902 device of the AMD/Xilinx VCK190 evaluation board, showing up to \textbf{5.44 TFLOPs} and \textbf{77.01 TOPs} throughput for fp32 and int8 precisions, respectively. MaxEVA significantly outperforms the state-of-the-art approach by presenting up to \textbf{2.19$\times$} higher performance and \textbf{20.4\%} energy efficiency gain.
  \item Open-sourcing MaxEVA for users to exploit our code in their designs, and contributing further to the knowledge of Versal AIE
  (\emph{\textbf{\href{https://github.com/enyac-group/MaxEVA}{https://github.com/enyac-group/MaxEVA}}}).
    
\end{itemize}











\section{Related Work}
\label{sec:Related_work}

The MatMul operation forms the core computation in many FPGA-based Deep Neural Network (DNN) accelerators.
For example, Sextans \cite{Sextans_FPGA22} is a general purpose MatMul accelerator evaluated on AMD/Xilinx U280 HBM FPGA.
Another work is \cite{Eriko_FPGAvsGPU_FPGA17}, where the authors implement MatMul-based accelerators for sparse, binary and ternary DNNs on Intel Arria 10 and Stratix 10 FPGAs.
In \cite{HARPv2_FPGA18} the authors present a multi-precision acceleration framework targeting the Intel HARPv2 CPU+FPGA platform, while in \cite{Boutros_beyond_FPT20} a MatMul accelerator is designed utilizing Intel's AI tensor blocks \cite{intel_tensor_block}.

Other works present MatMul FPGA accelerators optimized for specific DNN types, such as Convolutional Neural Networks (CNNs) \cite{Caffeine_CNN_FPGA19, Opt_CNN_FPGA_TCS20, CNN_FPGA_HBM_FCCM21}, and Transformers \cite{Transformer_ISLPED20, Transformer_ISQED21, Transformer_TECS22}.
Additionally, some works include automated frameworks for generating MatMul accelerators on FPGAs \cite{AutoSA_FPGA21, Auto_DNN_FCCM17, PolySA_ICCAD18, Auto_systolic_DAC17}, while others propose OpenCL FPGA accelerators 
\cite{OpenCL_FPT17, OpenCL_Arria_FPGA17}.


Although Versal ACAP is a new architecture, there exist several works that make use of AIEs in various application domains.
For instance, CHARM \cite{charm2023fpga} proposes multiple diverse MatMul accelerators on AIEs utilizing the VCK190 and achieving up to 2.94 TFLOPs for DNN inference.
In an extension of their work \cite{dac23automm}, the authors propose a framework to systematically generate MatMul accelerators on Versal AIE.
Their experiments on VCK190 device show higher energy efficiency, up to 1.7$\times$, compared to GPUs.
Other works on AIEs include CNN accelerators \cite{XVDPU_AIE_FPL22}, \cite{CNN_acc_Versal_FPL22}, as well as 
Graph Neural Network (GNN) acceleration \cite{H_GCN_FPL2022}.
Vyasa \cite{AIE_compiler_HPEC20} is a vectorizing compiler which extends the Halide DSL compiler \cite{Halide_2013} to automatically generate code for Versal AIE.
Finally, some works target AIE acceleration in the application domain of atmospheric simulations and weather predictions \cite{sparta_ICS23, Stensil_AIE_FPGA23}.

Among all prior works, the frameworks presented in \cite{charm2023fpga, dac23automm} are the closest to our work.
Both works use the same accelerator architecture to map MatMul workloads on VCK190.
In this work, we show the superiority of the MaxEVA framework by comparing with the aforementioned state-of-the-art implementations.
In particular, MaxEVA achieves notable performance gains of \textbf{2.19$\times$} and \textbf{20.8\%} for int8 and fp32, respectively, as well as \textbf{20.4\%} higher energy efficiency for fp32, over the state-of-the-art designs.
The MaxEVA framework optimizes the MatMul mapping on Versal AIE, while avoiding the performance bottlenecks that prior works encounter.



\section{Versal AI Engine Architecture}
\label{sec:Versal_AIE_architecture}

In this section, we provide an overview of the AMD/Xilinx Versal AIE architecture, as well as its main data movement and communication mechanisms. 

\subsection{AI Engine Architecture}
\label{subsec:AIE_architecture}

\begin{figure}[htbp]
\centering
\includegraphics[width=0.89\linewidth]{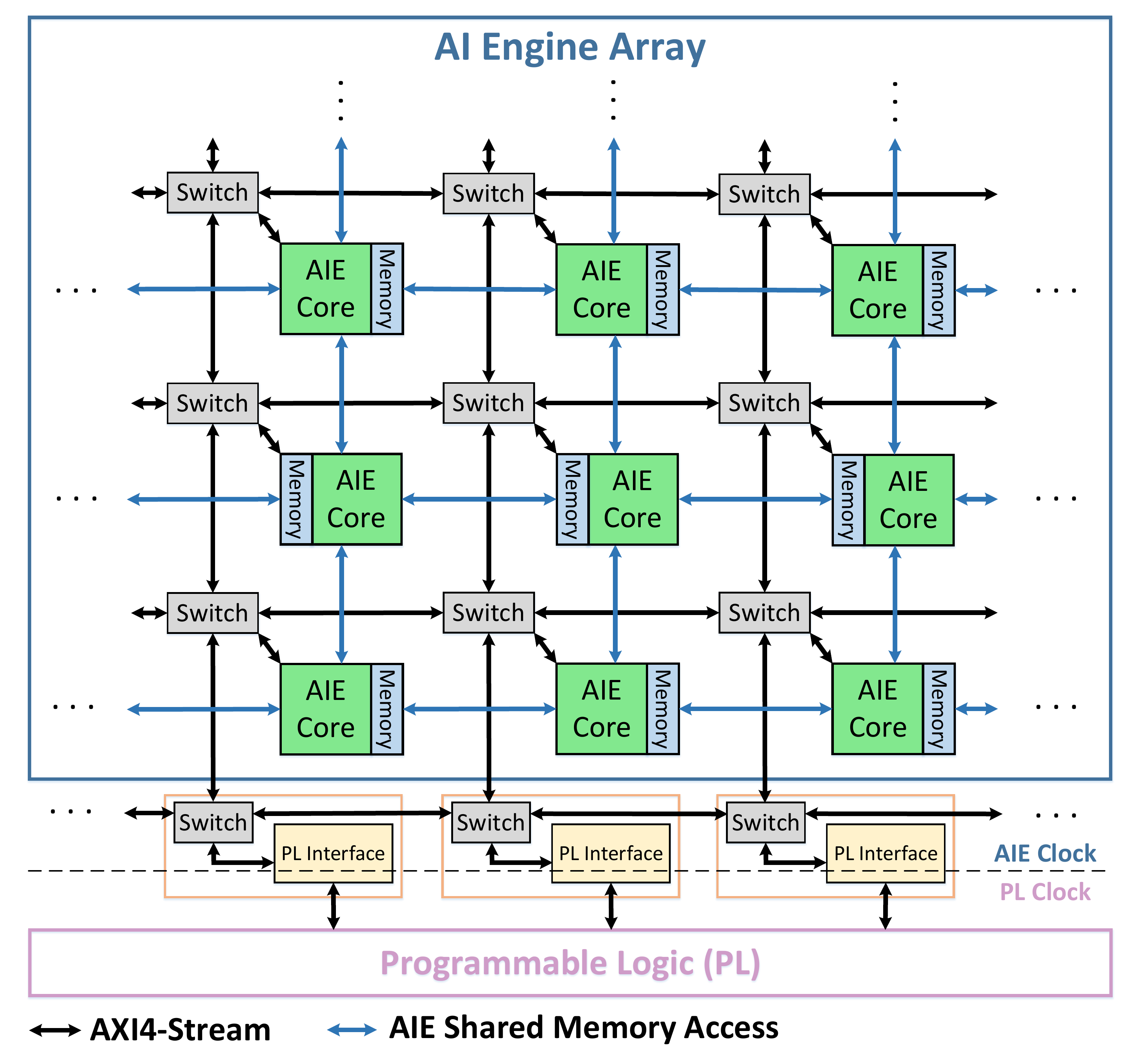}
\caption{Versal AI Engine architecture.}
\label{fig:versal_AIE_architecture}
\vspace{-0.38cm}
\end{figure}

The Versal AIE architecture is illustrated in Fig. \ref{fig:versal_AIE_architecture}.
The AIE architecture comprises a 2D array of homogeneous AIE tiles, where each tile contains an AIE core, a memory module, as well as an interconnection module (switch) \cite{Versal_acap_architecture_manual}.
The AIE array supports effectively 
three levels of parallelism:
first, each AIE core contains a Single-Instruction Multiple-Data (SIMD) vector processor, which allows multiple elements to be computed in parallel (data-level parallelism).
Second, the AIE core is architectured as a 7-way Very-Long Instruction Word (VLIW) processor, enabling multiple instructions to be executed every clock cycle (instruction-level parallelism).
Third, multiple AIE cores are able to execute in parallel across the AIE array (spatial parallelism).



Besides the SIMD processor, each AIE core also includes a scalar processing unit.
Both processors support fixed-point and floating-point precisions.
The AIE cores can be programmed by either using high-level C/C++ code utilizing the AIE API \cite{AIE_API_user_guide} or low-level SIMD intrinsics \cite{AI_Engine_programming_guide}.
To 
map an application to multiple AIE cores, AMD/Xilinx provides an Adaptive Data Flow (ADF) graph-based modelling.
The nodes in the ADF represent compute kernel functions and/or sub-graphs, while the edges represent the data connections among them \cite{AI_Engine_programming_guide}.
The data connections between AIE cores are realized through either direct memory sharing for neighboring AIEs or the AXI4-Stream 
switches for distantly located cores (Fig. \ref{fig:versal_AIE_architecture}).




In addition to AIE array, the Versal architecture combines the Processing System (PS), as well as the Programmable Logic (PL), all on the same chip. 
The PS consists of ARM CPUs, while the PL comprises the traditional FPGA resources, such as Look-Up Tables (LUTs), Block RAMs (BRAMs) and Digital Signal Processors (DSPs).
The communication of the AIE array with the other parts of the Versal device is realized through interface tiles, located on the last row of the array, as depicted in Fig. \ref{fig:versal_AIE_architecture}.
There are two types of interface tiles; the AIE-PL tiles and the AIE-NOC tiles.
The former provide dedicated connections with the PL, while the latter allow flexible communication with the other parts of the Versal chip through a Network-on-Chip (NOC) connection (not shown in Fig. \ref{fig:versal_AIE_architecture}).
The dedicated AIE-PL interface tiles contain primarily a PL interface which supports two different clock domains, \textit{i.e.}, the AIE clock and the PL clock, along with an AXI4-Stream switch to enable higher connection flexibility \cite{Versal_acap_architecture_manual}.

\begin{figure}[tbp]
\centering
\includegraphics[width=0.79\linewidth]{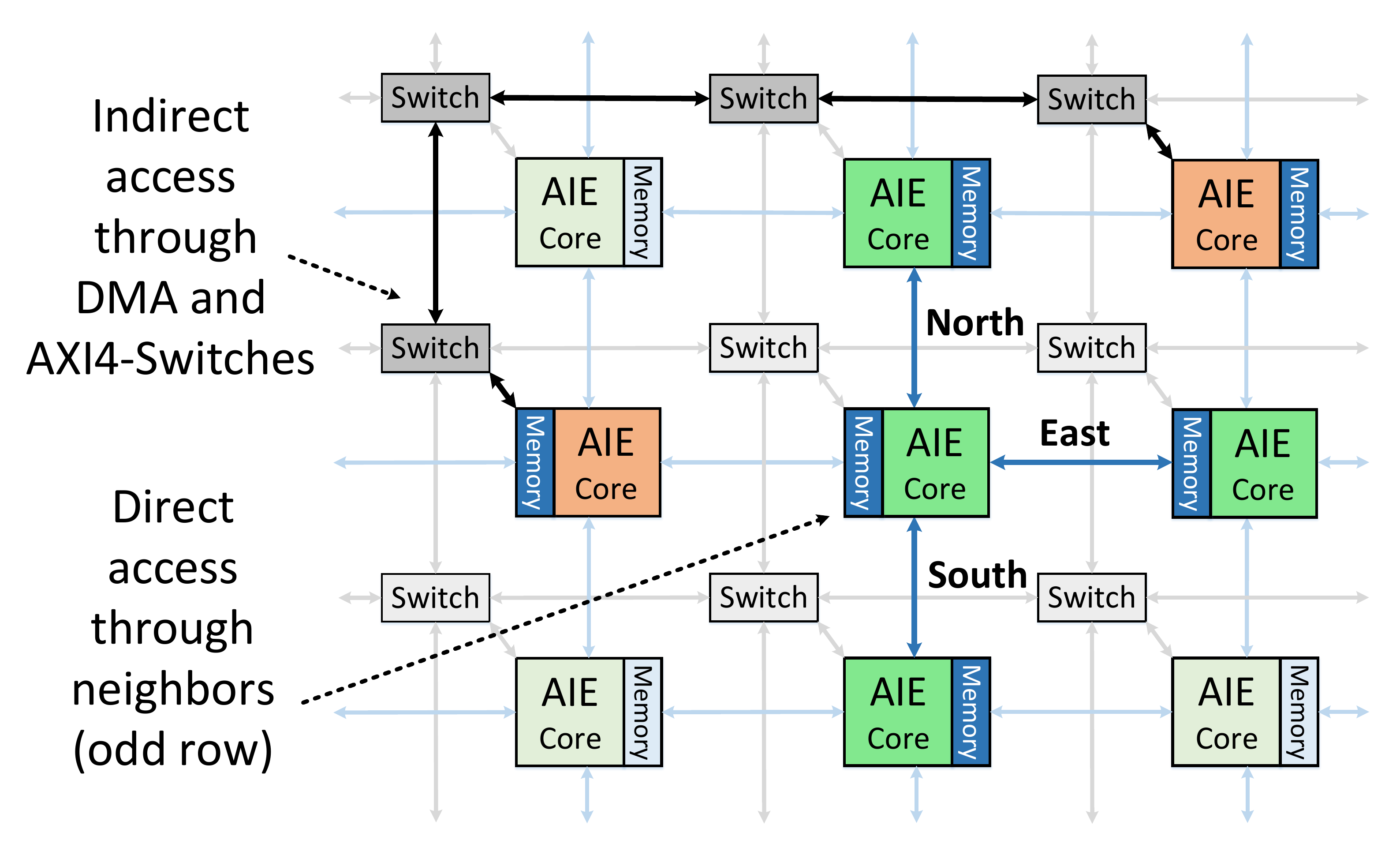}
\caption{Data movement mechanisms in AI Engine array.}
\label{fig:Data_movement_mechanims}
\vspace{-0.40cm}
\end{figure}

\subsection{AI Engine Data Movement Mechanisms}
\label{subsec:AIE_data_movement_mechanism}
Each AIE tile has 16KB of program memory to store VLIW instructions, as well as 32KB of data memory divided into 8 banks of 4KB.
For higher memory requirements, AIEs can access data memory directly from their neighbors, for a total of 128KB.
Fig. \ref{fig:Data_movement_mechanims} shows this direct access (AIEs highlighted in green), which constitutes the main data movement mechanism of the AIE array \cite{Versal_acap_architecture_manual}.
However, notice that while each AIE is able to directly access memory from its north and south directions, the east and west access depend on the row location of the AIE.
In particular, the AIE array is arranged on alternate even and odd rows, where the cores in even rows can access memory on the west direction, while in odd rows, the east module is accessed.
Finally, we note 
that AIEs on the edges of the array have fewer memory accesses on both north/south and east/west directions following the pattern described above.


For non-neighboring AIEs, the communication is realized through the Direct Memory Access (DMA) mechanism using the programmable switches, as shown in Fig. \ref{fig:Data_movement_mechanims} (AIEs highlighted in orange).
Compared to direct access, non-neighboring communication through DMA has increased communication latency and requires more memory resources.
The AXI4-Stream switches can be configured for either circuit-switching or packet-switching.
Circuit-switching provides dedicated connections which are  statically configured at compilation time.
In contrast, packet-switching allows routing to different destinations by dynamically setting a destination header at the beginning of each packet.
Due to static configuration, circuit-switching exhibits deterministic latency between connections, while also supporting broadcast to multiple output channels.
Conversely, packet-switching can cause resource contention, leading to non-deterministic latency \cite{Versal_acap_architecture_manual}.
In this work, we only exploit the most efficient circuit-switching mechanism, without the need of explicit packet-switching proposed in \cite{charm2023fpga, dac23automm}, as we discuss in the following Section.



\begin{figure}[htbp]
\centering
\includegraphics[width=0.70\linewidth]{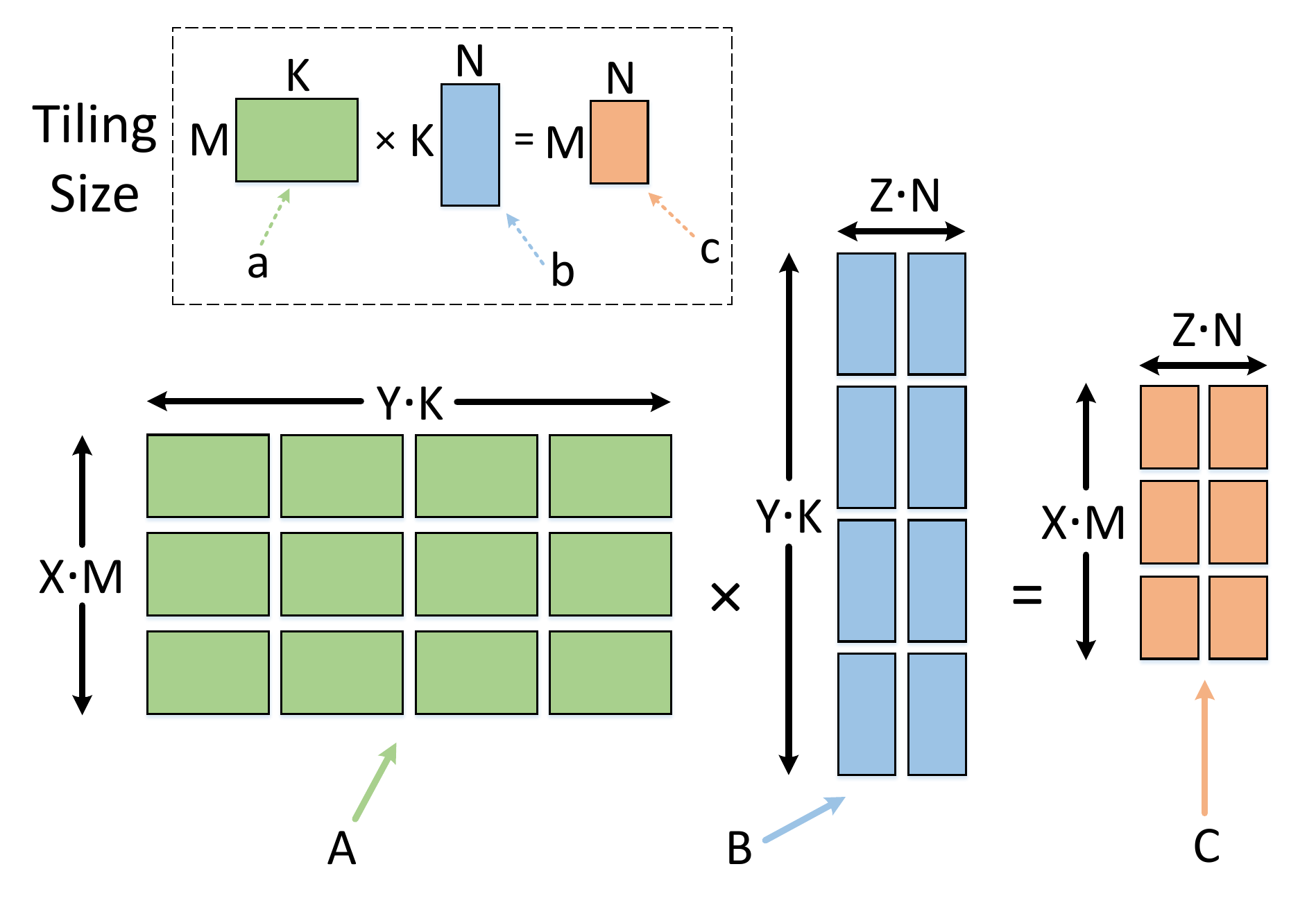}
\caption{Simplified view of tiling scheme for X=3, Y=4, Z=2.}
\label{fig:tiling_scheme}
\vspace{-0.40cm}
\end{figure}

\section{MaxEVA Framework}
\label{sec:Framework}

In this section we discuss the details of the MaxEVA framework.
The MaxEVA framework addresses the design, mapping and optimization of MatMul 
on the AIE array.
MaxEVA assumes that input/output data buffers are placed in PL BRAMs, as repeatedly used in practice to efficiently exploit data reuse in large matrices \cite{charm2023fpga, AutoSA_FPGA21, dac23automm, Caffeine_CNN_FPGA19}.
Through optimization based on analytical modelling and sophisticated kernel placement techniques, MaxEVA maximizes the throughput and energy efficiency of MatMul workloads on Versal AIE.


In this work, we focus on VC1902 device of the VCK190 board \cite{VCK_190_manual}, which has a total of 400 AIEs organized as an 
8 rows $\times$ 50 columns array.
As described in Section \ref{subsec:AIE_architecture}, the AIE/PL communication is established through the AIE-PL interface tiles.
However, not all existing columns in the AIE array can interface with PL.
For instance, in VC1902 there are only 
39 AIE-PL tiles \cite{Versal_AI_DC_AC_switching}.
The small AIE-PL bandwidth is one of the main challenges when designing MatMul applications on AIEs, which MaxEVA successfully overcomes.
Finally, 
although we show our method 
on the VC1902, our work can be 
generalized in straightforward fashion to any Versal device.


\subsection{Matrix Multiplication Tiling Scheme}
\label{subsec:tiling_scheme}

Fig. \ref{fig:tiling_scheme} depicts a simplified example of our proposed tiling scheme.
In our design, the tiling size ($M$$\times$$K$$\times$$N$) 
is determined by the single MatMul kernel, which is mapped to exactly one AIE core.
Since the Versal AIE comprises multiple cores, we map multiple MatMul kernels on the AIE array
(described by the parameters $X, Y, Z$ as explained below).
With this scheme, the final MatMul size running on the entire AIE array is 
$(X \cdot M) \times (Y \cdot K) \times (Z \cdot N)$. 
To this end, $X, Y, Z, M, K, N$ constitute the configurable integer parameters which are optimized by the MaxEVA framework.

\subsection{Matrix Multiplication Mapping on AI Engine Array}
\label{subsec:MatMul_mapping_on_AIE}

\begin{figure}[htbp]
\centering
\includegraphics[width=0.90\linewidth]{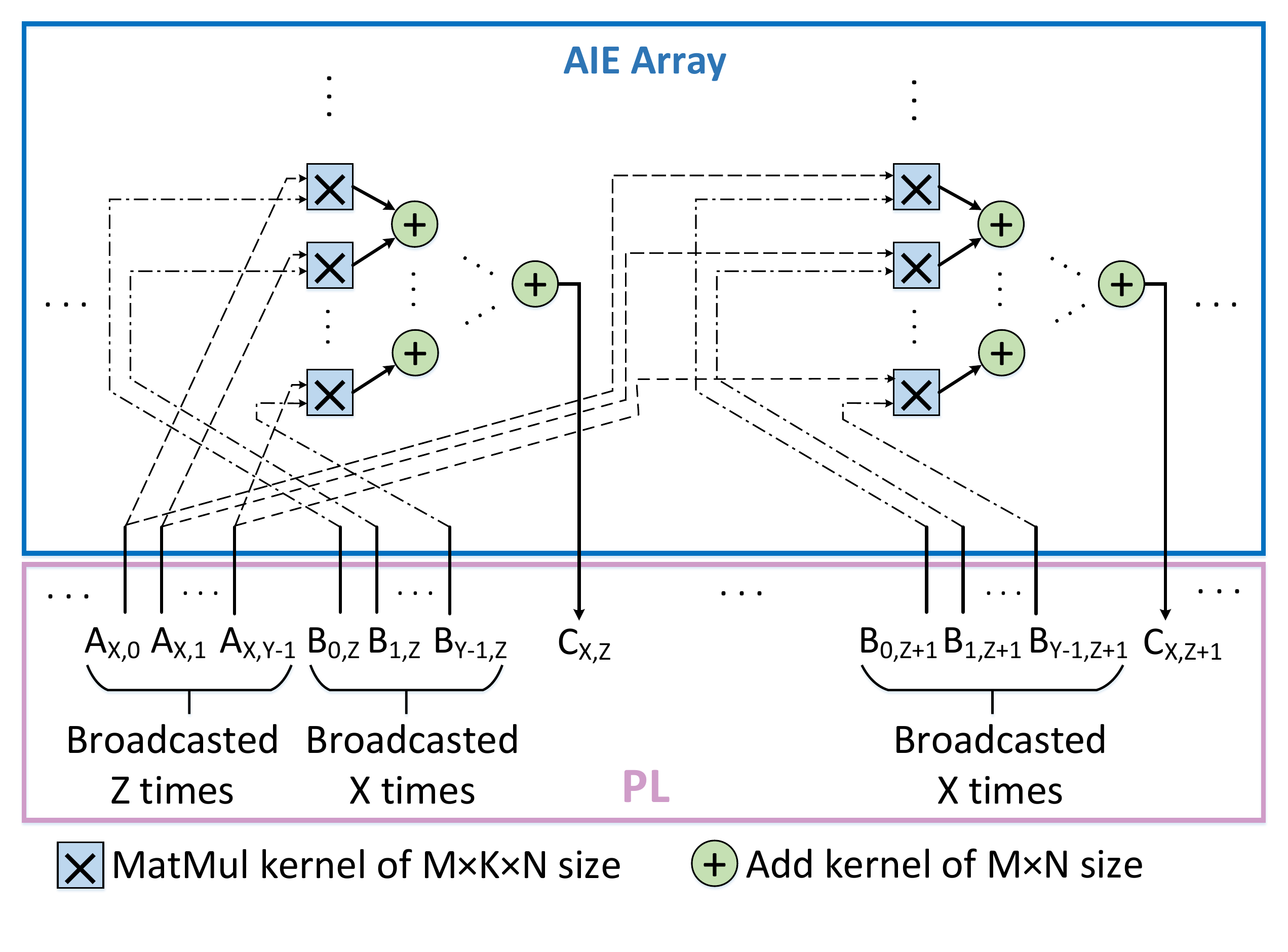}
\caption{MatMul mapping on AIE array by leveraging input broadcasting and output adder tree reduction.}
\label{fig:MatMul_mapping_on_AIE_array}
\vspace{-0.40cm}
\end{figure}

To overcome the reduced number of AIE-PL interface tiles, and thus avoid the PL Input/Output (PLIO) bottleneck, we leverage the two principal properties of the MatMul algorithm.
First, we exploit the inherent data reuse of the MatMul algorithm to reduce the number of incoming PLIOs (inputs to AIE array), by broadcasting inputs $A$ and $B$ (Fig. \ref{fig:tiling_scheme}) to multiple AIEs.
Second, we utilize the native reduction of the $Y \cdot K$ dimension in MatMul to decrease the number of outgoing PLIOs (outputs of AIE array) by performing reduction on the AIE itself, instead of the PL.
With this method, we are able to efficiently map $X \cdot Y\cdot Z$ MatMul kernels 
(each performing a MatMul computation of $M$$\times$$K$$\times$$N$ size), and $X \cdot (Y-1)\cdot Z$ Add kernels (each reducing partial results of $M$$\times$$N$ size).

Fig. \ref{fig:MatMul_mapping_on_AIE_array} shows a high-level mapping diagram of MatMul and Add kernels on the AIE array.
In our design, there exist groups of $Y$ MatMul kernels along with their corresponding adder trees ($Y-1$ adders on each group).
More specifically, there are $X \cdot Z$ of such groups in total, all executing in a parallel fashion.
Each $A_{x,y}$ and $B_{y,z}$ PLIO input data are broadcasted to their corresponding MatMul kernels, $Z$ and $X$ times, respectively, as governed by the MatMul algorithm.
In our design, there are in total $X \cdot Y + Y \cdot Z$ PLIO inputs, as well as $X \cdot Z$ PLIO outputs.
We note here that broadcasts are realized through the programmable switches of the AIE array and are statically configured during compilation (circuit-switching).

\begin{figure}[tbp]
\centering
\includegraphics[width=0.60\linewidth]{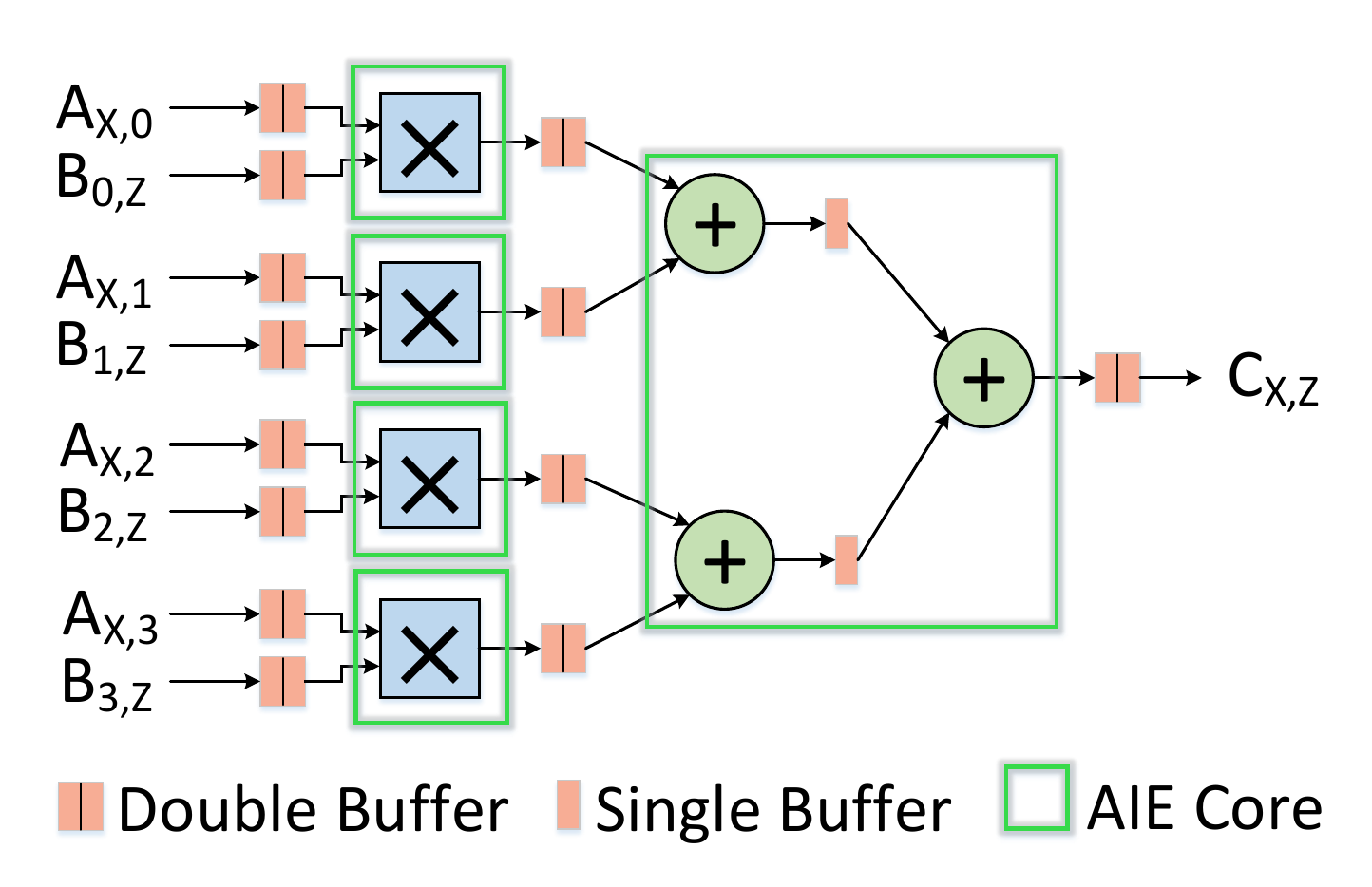}
\caption{Group of MatMul kernels (each mapped to one AIE core) and its adder tree (whole tree mapped to one AIE core).}
\label{fig:adder_tree_mults_double_single_buffers}
\vspace{-0.40cm}
\end{figure}

Fig. \ref{fig:adder_tree_mults_double_single_buffers} illustrates a group comprising 4 MatMul kernels along with its adder tree (3 Add kernels).
As mentioned above, each MatMul kernel executes on a separate AIE core.
However, we map the whole adder tree to a single AIE core, where all Add kernels execute in a sequential fashion.
We make such design choice for various reasons.
First, we note that only MatMul kernels contribute to throughput, while Add kernels are only used to reduce the output PLIOs.
Thus, we maximize the number of MatMul kernels, by minimizing the AIE cores used to run the Add kernels.
Second, Add kernel's execution time (latency) is much lower than MatMul kernel's latency.
Therefore, we are able to map multiple Add kernels to a single AIE core, without any performance degradation (as we show in Section \ref{subsec:Single_kernel_results}).
Third, when mapping multiple kernels to a single AIE core, memory resources are reduced compared to mapping to several AIE cores.
As depicted in Fig. \ref{fig:adder_tree_mults_double_single_buffers}, double buffers are inserted between separate AIE cores to effectively overlap computation with communication (to increase the compute utilization of AIEs).
However, between the Add kernels only single buffers are used, since all Add kernels are executed sequentially.
This results in twofold memory buffer savings compared to if they were executed on separate AIEs.
Overall, the throughput of the entire design is determined by the computationally heavy MatMul kernels, since the whole adder tree latency is lower than MatMul latency (Section \ref{subsec:Single_kernel_results}).

\subsection{AI Engine Kernels Modelling \& Optimization}
\label{subsec:AIE_kernels_optimization}

Since the design space of the configurable integer parameters $X, Y, Z, M, K, N$ is large, we propose an analytical model to maximize the throughput of MatMul on the AIE array.
Our model takes as input device-specific parameters and constraints, \textit{e.g.}, I/O bandwidth, AIE peak throughput, number of PLIOs, and finds the optimal solution based on our mapping methodology.
In this section, first, we describe the optimization of the single MatMul kernel, and then we discuss mapping the entire MatMul to the AIE array.


1) \textit{Single AI Engine Kernel Optimization: M, K, N parameters}.
Our model takes as input a lower bound of the  efficiency of the single MatMul kernel, $eff\_lb$,
to ensure that the achieved throughput is very close to the theoretical AIE throughput.
Here, we define efficiency ($eff$) as the fraction of the achieved throughput to the peak throughput of the vector processor inside the AIE core.
In particular, each AIE vector processor is able to achieve up to 128 MACs/cyc (multiply-accumulate operations per clock cycle) for int8 precision, while for fp32 the peak throughput is 8 MACs/cyc \cite{Versal_acap_architecture_manual}. 
Defining $peak\_MACs$ as the AIE core peak throughput and $kernel\_cyc$ as the latency (in clock cycles) of the MatMul kernel of $M$$\times$$K$$\times$$N$ size, we get the following constraint: 
\begin{gather} 
eff \geq eff\_lb \Rightarrow \nonumber \\
\left(\frac{M \cdot K \cdot N}{kernel\_cyc}\right)/peak\_MACs \geq eff\_lb \Rightarrow \nonumber \\
kernel\_cyc \leq M \cdot K \cdot N / \left(eff\_lb \cdot peak\_MACs\right) \label{eq:kernel_throughput}
\end{gather}
Next, we optimize our design by ensuring that I/O bandwidth does not become a performance bottleneck.
There are two I/O bandwidth considerations in our design: the PLIO bandwidth for both inputs and outputs, as well as the bandwidth of the AXI4-Stream switches. 
According to \cite{Versal_acap_architecture_manual}, the bandwidth for both I/Os is $BW\_IO = 4Bytes/cyc$. 
To assure that our design is not I/O limited, we require the latency of input ($a\_cyc$, $b\_cyc$) and output ($c\_cyc$) data transmission to be lower than the MatMul kernel latency. Therefore, we get:
\begin{align}
\{a\_cyc, \ b\_cyc, \ c\_cyc\} \le & kernel\_cyc \Rightarrow \nonumber \\
\{ M \cdot K \cdot sizeof(a)/BW\_IO &\le kernel\_cyc ,\nonumber\\
K \cdot N \cdot sizeof(b)/BW\_IO &\le kernel\_cyc , \nonumber\\
M \cdot N \cdot sizeof(c)/BW\_IO &\le kernel\_cyc \} 
\label{eq:plio}
\end{align}

By combining equations \ref{eq:kernel_throughput} and \ref{eq:plio}, we have the following:
\begin{align}
N \geq eff\_lb \cdot peak\_MACs &\cdot sizeof(a) / BW\_IO \label{eq:a_BW_constr}\\
M \geq eff\_lb \cdot peak\_MACs &\cdot sizeof(b) / BW\_IO \label{eq:b_BW_constr}\\
K \geq eff\_lb \cdot peak\_MACs &\cdot sizeof(c) / BW\_IO \label{eq:c_BW_constr}
\end{align}
Notice that $a$, $b$ are the inputs of the MatMul kernel, while $c$ is the output of either the MatMul or the Add kernel (both have same output size of $M$$\times$$N$).
The $sizeof()$ function calculates the size (in Bytes) of input/output \textit{data types}.
This is particularly important for int8 computation, since we perform all accumulations in 32-bits (int32).
In this case, the size of inputs $a$, $b$ is 1 Byte, while the size of output $c$ is 4 Bytes.

Finally, we define a constraint that all input/output buffers of the single MatMul kernel should fit within the local AIE memory.
By not exceeding the local AIE memory, we are able to maximize the number of MatMul kernels that execute in parallel on the AIE array.
Each AIE core needs some system memory for its operation, \textit{e.g.}, stack, heap. The AIE data memory is partitioned in 4KB banks; we leave one bank for system memory. This implies an available space of 28KB for our/user buffers.
Because both input and output buffers of MatMul kernels are double buffered (see Fig. \ref{fig:adder_tree_mults_double_single_buffers}), we get:
\begin{multline}
\{M \cdot K \cdot sizeof(a) + K \cdot N \cdot sizeof(b)\\ 
+ M \cdot N \cdot sizeof(c)\} \le 14KB \label{eq:memory_constr}
\end{multline}

The solution of $M, K, N$ can be formulated as an integer programming (IP) optimization problem, where we maximize the number of MACs of the single MatMul kernel by having eq. \ref{eq:a_BW_constr}--\ref{eq:memory_constr} as constraints.
Increasing the number of MACs will lead to more data reuse in the vector registers of the AIE core, resulting in higher kernel efficiency.
The lower bound of the efficiency ($eff\_lb$) can be assigned based on the performance constraints of the application and the achievable throughput.
In this work, we are able to achieve 95\% of MatMul kernel efficiency (Section \ref{subsec:Single_kernel_results}), which we set it as our lower bound.
The configurable parameters are evaluated through exhaustive search and top-ranked design points are reported.
We note that solving the IP exhaustively is a suitable approach, since the search space is significantly reduced by considering only powers of two for $M, K, N$ (as discussed in Section \ref{subsec:Single_kernel_results}).


2) \textit{Multiple AI Engine Kernels Optimization: X, Y, Z parameters}.
To obtain an optimal mapping onto the AIE array, we require our entire design to fit in the total number of AIE cores ($AIE\_cores$).
We also require the number of utilized input/output PLIOs to not exceed the available PLIOs of the device.
In particular, for VC1902, $PLIO\_in = 78$ and $PLIO\_out = 117$ \cite{Versal_acap_architecture_manual, Versal_AI_DC_AC_switching}.
Based on the discussion at Section \ref{subsec:MatMul_mapping_on_AIE} and the above, the following can be expressed:
\begin{gather}
X \cdot Y \cdot Z + X \cdot Z \le AIE\_cores \label{eq:total_cores_constr} \\
X \cdot Y + Y \cdot Z \le PLIO\_in  \label{eq:PLIO_in_constr}\\
X \cdot Z \le PLIO\_out \label{eq:PLIO_out_constr}
\end{gather}

The optimization of $X, Y, Z$ is evaluated through exhaustive search by maximizing the total number of MatMul kernels ($X \cdot Y \cdot Z$), which leads to maximized throughput of the MatMul application.
Again, exhaustive search is sufficient because all constants in eq. \ref{eq:total_cores_constr}--\ref{eq:PLIO_out_constr} are in the order of hundreds (reasonably small) \cite{Versal_acap_architecture_manual}.
Multiple top-ranked design points are reported, from which we explore various options (refer to Section \ref{subsec:Multiple_kernels_results}).

\subsection{AI Engine Kernel Placement}
\label{subsec:Placement_strategy}

\begin{figure}[bp]
\vspace{-0.3cm}
\centering
\includegraphics[width=0.55\linewidth]{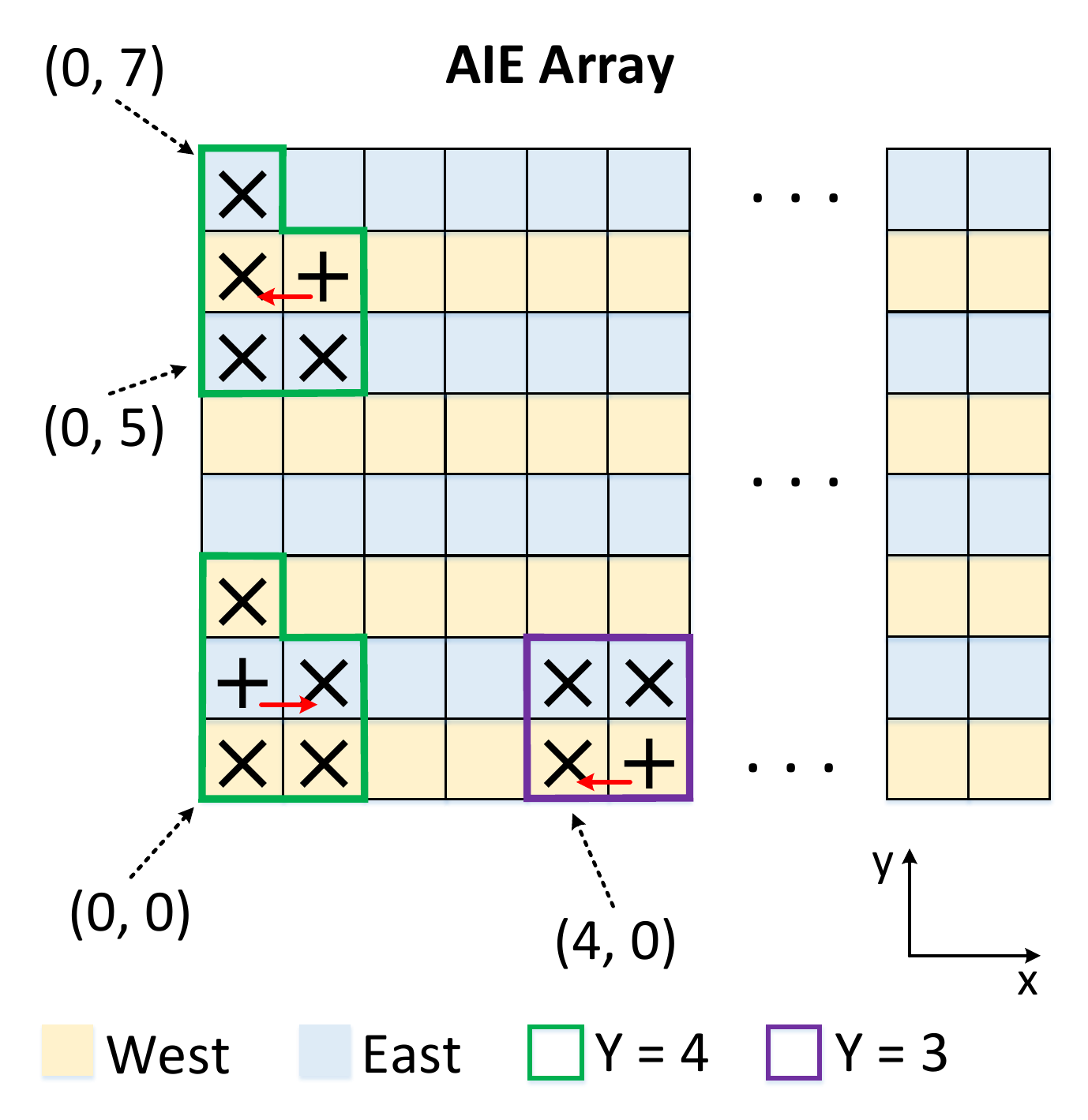}
\caption{Examples of proposed placement strategy for $Y=3, 4$.}
\label{fig:placement_example}
\end{figure}

To leverage the most efficient local data sharing mechanism described in Section \ref{subsec:AIE_data_movement_mechanism}, we propose a sophisticated kernel placement strategy.
Fig. \ref{fig:placement_example} illustrates an example of the placement procedure, where each \textit{multiplication} symbol denotes a MatMul kernel, while the \textit{addition} symbol indicates the adder tree mapped to a single AIE core.
We place each group of $Y$ MatMul kernels along with its adder tree, in a manner to avoid any DMA usage in the buffer connections \textit{between} MatMul and Add kernels (Fig. \ref{fig:adder_tree_mults_double_single_buffers}).
For instance, when considering the group starting at (0, 0) location in Fig. \ref{fig:placement_example}, the adder tree is able to access the memory of 3 (out of 4) MatMul kernels directly (along the north, south and east direction).
Notice that in this case, the MatMul kernel located at (1, 0) does not directly communicate with the adder tree.
However, the output buffer of this MatMul kernel can be placed to its north location (1, 1); this is possible because of direct memory sharing between neighboring AIEs as shown in Fig \ref{fig:Data_movement_mechanims}. From here the adder tree AIE can access it directly, thus ensuring no DMA usage.

\begin{figure}[htbp]
\centering
\includegraphics[width=0.95\linewidth]{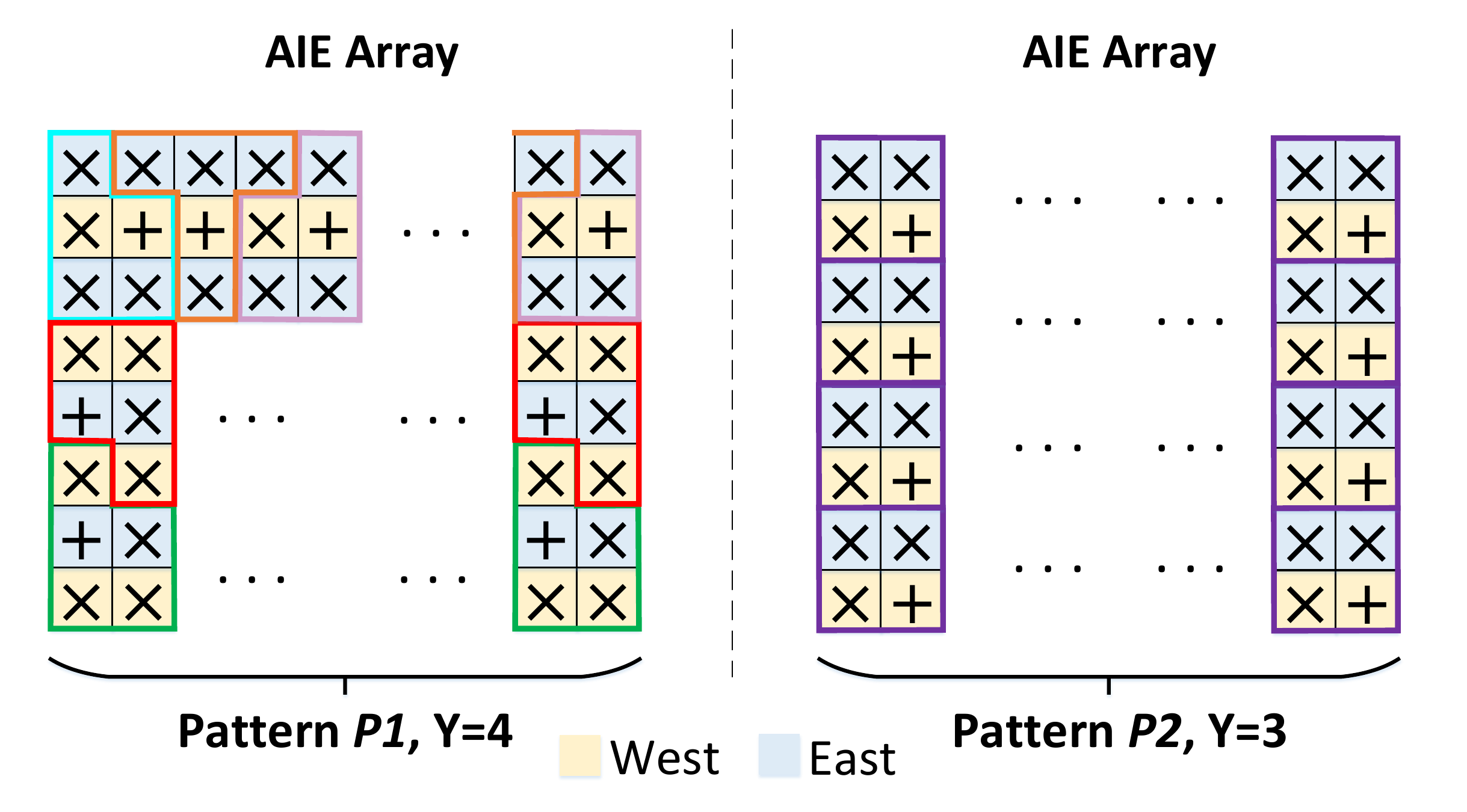}
\caption{Proposed placement patterns.}
\label{fig:placement_patterns}
\vspace{-0.38cm}
\end{figure}

Another placement example is the group starting at (0, 5).
Although, this placement is similar to the (0, 0) case described above, notice that the adder tree is located on the opposite side.
This is because the local data sharing is realized only on the west direction in even rows.
We observe that in this case, the adder tree can only access 2 out of 4 MatMul kernels directly (located at (0, 6) and (1, 5)). 
However, the output buffers of all MatMul kernels can be placed such that no DMA is used.
For instance, the output buffers of (0, 5) and (0, 7) kernels can be placed at (1, 5) and (1, 7) locations (east access), respectively, which can be both directly accessed by the adder tree.



An example for a group of $Y=3$ kernels is also shown in Fig. \ref{fig:placement_example}, where the explanation of its placement is similar to the examples discussed above.
Finally, we note that the input data buffers of the MatMul kernels, as well as the output data buffer of the adder tree (refer to Fig. \ref{fig:adder_tree_mults_double_single_buffers}) can be placed on any free memory space, again only accessing memory directly.
We let the AMD/Xilinx AIE PnR (place and route) tool to make such decisions and optimize the whole AIE array mapping.


We exploit the aforementioned placement strategy to fill the entire AIE array, thus mapping multiple MatMul/Add kernels.
Fig. \ref{fig:placement_patterns} shows our two proposed placement patterns, $P1$ and $P2$. 
Notice that in pattern $P1$, we use a new \emph{``T"--like} shape outlined in orange color, while all other shapes of both patterns are similar to the examples in Fig. \ref{fig:placement_example}.
This is required to fill the whole AIE array and ensure that no AIEs remain unutilized.
However, each \emph{``T"--like} shape will lead to a small DMA usage (one MatMul output buffer), as we show in the next Section.
Finally, we only propose patterns for $Y=3, 4$, because these constitute the top-ranked solutions based on our optimization approach, as we also present in the next Section.

\section{Evaluation}
\label{sec:Evaluation}

In this section, we first report single AIE kernel experimental results.
Second, we present results and evaluation of full  MatMul mapping on the AIE array.
Finally, we perform a comprehensive comparison with the state-of-the-art approach, which exhibits the superiority of the MaxEVA framework.

We compile and simulate our designs by using the AMD/Xilinx Vitis 2022.1 version.
Across all experiments, the AIE frequency is set to the maximum frequency of the VC1902 on VCK190, \textit{i.e.}, 1.25 GHz, while the PL operates at 312.5 MHz, as recommended in \cite{AI_Engine_programming_guide, AI_Engine_programming_environment}.
To ensure rate matching without performance reduction between AIE and PL, we set the PLIO width to 128-bits \cite{AI_Engine_programming_guide}.
Moreover, we use the AIE simulator \cite{AI_Engine_programming_environment}   
to calculate the throughput of our designs, which we report as an average of 10 runs.
Finally, power consumption is estimated through the AIE XPE tool \cite{AIE_XPE}.

Since our focus is to achieve maximum MatMul throughput from the AIE array, 
isolating the AIE implementation from any source of performance degradation is crucial to obtain accurate evaluation.
Such sources of performance reduction may arise from any design mapped to PL causing stalls, as well as the limited DRAM bandwidth.
Therefore, we simulate our AIE designs without utilizing the PL and DRAM.
We simulate the state-of-the-art CHARM framework as well by leveraging their open-source code \cite{charm2023fpga, dac23automm}, in the exactly same manner, with the same assumptions.
This ensures a thorough and fair comparison between MaxEVA and CHARM frameworks, thus enabling us to report accurate throughput, power, and AIE resource utilization.
However, we note that the CHARM code includes only fp32 implementation.
In \cite{dac23automm} the authors report results for int8 implementation as well, but their code is not publicly available.
Therefore, for int8, we show a qualitative comparison based on their reported results.

\begin{table}[t]
 \centering
\caption{Single AI Engine kernel results.}
\resizebox{1.00\linewidth}{!}{
\begin{tabular}{c|c|c|c|c}
\Xhline{2.5\arrayrulewidth}
\textbf{Kernel}  & \textbf{Kernel size}  & \textbf{Latency}  & \textbf{Throughput}  & \textbf{Effic-} \\

\textbf{type} & \textbf{M$\times$K$\times$N} & \textbf{(cyc)} & \textbf{(MACs/cyc)} & \textbf{iency} \\
\hline
\hline

MatMul int8 & 32$\times$128$\times$32 & 1075 (1$\times$) & 121.93 & 95.26\% \\
Add int32 & 32$\times$32 & 164 (0.15$\times$) & 6.24 & 78.05\% \\
\hline
MatMul fp32 \cite{charm2023fpga, dac23automm} & 32$\times$32$\times$32 & 4329 (1$\times$) & 7.57 & 94.70\% \\
Add fp32 & 32$\times$32 & 167 (0.04$\times$) & 6.13 & 76.65\% \\

\Xhline{2.5\arrayrulewidth}

\end{tabular}

}
\label{tb:single_AI_engine_results}
\vspace{-0.18cm}
\end{table}

\subsection{Performance of Single AI Engine Kernels}
\label{subsec:Single_kernel_results}

For maximum efficiency, MatMul and Add kernels have been designed to leverage the vector processors of the AIEs.
Our kernels are written in C/C++ utilizing the AIE APIs and several optimization compiler directives (pragmas) that perform software pipelining, loop unrolling/flattening and explicit independence between data \cite{AI_Engine_programming_guide}.
Moreover, during our kernel design experimentation, we found that MatMul kernels with powers of two dimensions produce higher efficiency.
Hence, in this work, we use powers of two during the optimization of the $M, K, N$ parameters, as also proposed in \cite{charm2023fpga, H_GCN_FPL2022}.

\begin{table*}[ht]
 \centering
\caption{Evaluation of several MaxEVA configurations for fp32 and comparison with state-of-the-art approach.}

\resizebox{1.00\textwidth}{!}{
\begin{tabular}{c|c|c|c|c|c|c|c|c||c|c}
\Xhline{2.5\arrayrulewidth}
\textbf{MaxEVA Cfg.}  & \textbf{MatMul}  & \textbf{Total}  & \textbf{Memory} & \textbf{DMA}  & \textbf{PLIOs}  & \textbf{Throughput} & \textbf{Power} & \textbf{Energy Eff.} & \textbf{AIE core} & \textbf{Memory}\\

\textbf{X$\times$Y$\times$Z (pat.)} & \textbf{kernels} & \textbf{AIE cores} & \textbf{banks} & \textbf{banks} & \textbf{} & \textbf{(GFLOPs)} & \textbf{(W)} & \textbf{(GFLOPs/W)} & \textbf{P. (W)} & \textbf{P. (W)}\\
\hline
\hline

1. \hspace{1.5mm} 13$\times$4$\times$6 (P1) & 312 & 390 (97.5\%) & 3138 (98.1\%) & 18 & 154 (79.0\%) & \textbf{5442.11 (+20.8\%)} &  43.83 & \textbf{124.16 (+20.4\%)} & 25.62 & 18.21 \\
2. \hspace{0.2mm} 10$\times$3$\times$10 (P2) & 300 & 400 (100\%) & 3190 (99.7\%) & 0 & 160 (82.1\%) & 5405.33 (+20.0\%) & 44.66 & 121.03 (+17.4\%) & 25.54 &  19.12\\
3. \hspace{1.5mm} 11$\times$4$\times$7 (P1) & 308 & 385 (96.3\%) & 3106 (97.1\%) & 18 & 149 (76.4\%) & 5414.39 (+20.2\%) & 44.01 & 123.03 (+19.3\%) & 25.36 & 18.65 \\
4. \hspace{1.5mm} 11$\times$3$\times$9 (P2) & 297 & 396 (99.0\%) & 3176 (99.3\%) & 0 & 159 (81.5\%) & 5382.27 (+19.5\%) & 44.13 & 121.96 (+18.3\%) & 25.35 & 18.78 \\
5. \hspace{1.5mm} 12$\times$4$\times$6 (P1) & 288 & 360 (90.0\%) & 2934 (91.7\%) & 16 & 144 (73.8\%) & 5031.19 (+11.7\%) & 40.68 & 123.68 (+20.0\%) & 23.77 & 16.91 \\
6. \hspace{1.5mm} 12$\times$3$\times$8 (P2) & 288 & 384 (96.0\%) & 3092 (96.6\%) & 0 & 156 (80.0\%) & 5225.05 (+16.0\%) & 42.28 & 123.58 (+19.9\%) & 24.68 & 17.60 \\
\hline
\hline

CHARM \cite{charm2023fpga, dac23automm} & 384 & 384 (96.0\%) & 3086 (96.4\%) & 0 & 80 (41.0\%) & 4504.46 (+0\%) & 43.69 & 103.10 (+0\%) & 26.95 & 16.74 \\

\Xhline{2.5\arrayrulewidth}

\end{tabular}
}
\label{tb:MaxEVA_results_fp32}
\end{table*}

\begin{table*}[ht]
 \centering
\caption{Evaluation of several MaxEVA configurations for int8 (results for CHARM obtained from \cite{dac23automm}).}

\resizebox{0.96\textwidth}{!}{
\begin{tabular}{c|c|c|c|c|c|c|c|c||c|c}
\Xhline{2.5\arrayrulewidth}
\textbf{MaxEVA Cfg.}  & \textbf{MatMul}  & \textbf{Total}  & \textbf{Memory} & \textbf{DMA}  & \textbf{PLIOs}  & \textbf{Throughput} & \textbf{Power} & \textbf{Energy Eff.} & \textbf{AIE core} & \textbf{Memory}\\

\textbf{X$\times$Y$\times$Z (pat.)} & \textbf{kernels} & \textbf{AIE cores} & \textbf{banks} & \textbf{banks} & \textbf{} & \textbf{(TOPs)} & \textbf{(W)} & \textbf{(TOPs/W)} & \textbf{P. (W)} & \textbf{P. (W)}\\
\hline
\hline

1. \hspace{1.5mm} 13$\times$4$\times$6 (P1) & 312 & 390 (97.5\%) &  3112 (97.3\%) & 18 & 154 (79.0\%) & \textbf{77.01 (2.19$\times$)}& 66.83 & 1.152 & 48.65 & 18.18 \\
2. \hspace{0.2mm} 10$\times$3$\times$10 (P2) & 300 & 400 (100\%) &  3194 (99.8\%) & 0 & 160 (82.1\%) & 76.08 (2.16$\times$) & 65.52 & \textbf{1.161} & 47.44 & 19.08 \\
3. \hspace{1.5mm} 11$\times$4$\times$7 (P1) & 308 & 385 (96.3\%) &  3096 (96.8\%) & 18 & 149 (76.4\%) & 75.67 (2.15$\times$) & 66.79 & 1.133 & 48.17 & 18.62 \\
4. \hspace{1.5mm} 11$\times$3$\times$9 (P2) & 297 & 396 (99.0\%) &  3178 (99.3\%) & 0 & 159 (81.5\%) & 74.66 (2.12$\times$) & 65.83 & 1.134 & 47.04 & 18.79 \\
5. \hspace{1.5mm} 12$\times$4$\times$6 (P1) & 288 & 360 (90.0\%) &  2918 (91.2\%) & 16 & 144 (73.8\%) & 71.25 (2.02$\times$) & 62.13 & 1.147 & 45.15 & 16.98 \\
6. \hspace{1.5mm} 12$\times$3$\times$8 (P2) & 288 & 384 (96.0\%) &  3080 (96.3\%) & 0 & 156 (80.0\%) & 72.93 (2.07$\times$) & 63.24 & 1.153 & 45.71 & 17.53 \\
\hline
\hline

CHARM \cite{charm2023fpga, dac23automm} & 192 & 192 (48.0\%) & -- & -- & -- & 35.19 (1$\times$) & -- & -- & -- & -- \\

\Xhline{2.5\arrayrulewidth}

\end{tabular}
}
\label{tb:MaxEVA_results_int8}
\vspace{-0.30cm}
\end{table*}

Table \ref{tb:single_AI_engine_results} presents the single AIE kernel results.
For int8 precision, the 32$\times$128$\times$32 MatMul kernel was the only solution that satisfied all constraints in our optimization procedure (Section \ref{subsec:AIE_kernels_optimization}).
All other values for $M, K, N$ are either I/O bandwidth limited or exceed the memory constraint of 14KB (eq. \ref{eq:memory_constr}).
In contrast, for fp32, there are many top-ranked solutions that maximize the number of MACs, \textit{e.g.}, 16$\times$64$\times$32, 64$\times$16$\times$32, 32$\times$32$\times$32, \textit{etc.}
However, we notice that all best solutions exhibit the same number of MACs (equal to 32768).
Since 32$\times$32$\times$32 MatMul kernel is one of our optimized solutions and also used in state-of-the-art CHARM \cite{charm2023fpga, dac23automm}, we obtain its code from their open-source code base.
This ensures a fair comparison between their approach and ours. 

From Table \ref{tb:single_AI_engine_results}, we observe that the int8 MatMul kernel utilizes the vector processor of the AIE efficiently - a very high efficiency of 95.26\% is achieved.
The fp32 MatMul kernel obtained from \cite{charm2023fpga, dac23automm} also presents very high efficiency (94.70\%), and is designed by using AIE intrinsics.
Moreover, we observe that int8 and fp32  Add kernels have very similar latencies, which are both significantly lower than MatMul latencies (164 \textit{vs.} 1075 cycles for int8, and 167 \textit{vs.} 4329 cycles for fp32). 
This validates that multiple Add kernels are able to run sequentially into a single AIE, without causing any performance degradation.
We also observe that the relative latency ratio of Add kernel to MatMul kernel is notably lower for fp32 (0.04$\times$) compared to int8 (0.15$\times$).
These relative ratios indicate that the AIE core running even multiple Add kernels remains idle for substantially longer for the fp32 case, affecting its power consumption accordingly (Section \ref{subsec:Multiple_kernels_results}).
Finally, we note that Add kernels also exhibit high efficiency (78.05\% and 76.65\% for int8 and fp32, respectively), though not as high as the efficiency of MatMul kernels.
This performance difference is due to less data reuse on the AIE vector registers by Add kernels compared to MatMul kernels. 



\subsection{Performance of Matrix Multiplication on AI Engine Array}
\label{subsec:Multiple_kernels_results}

We utilize the MaxEVA framework to optimize the performance of the entire MatMul application.
To map multiple kernels to the AIE array we wrote a parameterized C++ code for any values of $X, Y, Z$ by exploiting the ADF graph model.



\subsubsection{MaxEVA \textit{vs.} state-of-the-art CHARM for fp32 precision}
From our multiple AIEs optimization of $X$$\times$$Y$$\times$$Z$ parameters, we found that the 10$\times$4$\times$8 solution maximizes the number of MatMul kernels. 
In this case, there are 320 MatMul kernels and 80 cores which run Add kernels, hence, all 400 AIE cores are utilized.
However, this solution was not feasible because the AIE PnR tool failed due to routing congestion.
This is due to the extra routing needed because of DMA usage (pattern $P1$), as well as the 100\% utilization of the AIE cores, leaving no free space for successful routing.
Our second top-ranked solution, \textit{i.e.}, 13$\times$4$\times$6, does not present any routing issues and is successfully mapped to the AIE array.
In Table \ref{tb:MaxEVA_results_fp32} we show this solution (row 1), which achieves a very high throughput of 5442.11 GFLOPs, outperforming the state-of-the-art approach by \textbf{20.8\%} 
(CHARM presents 4504.46 GFLOPs).

When further comparing the 13$\times$4$\times$6 MaxEVA solution to CHARM, we can observe from Table \ref{tb:MaxEVA_results_fp32} (row 1) that our method utilizes the AIE array slightly more (390 \textit{vs.} 384 AIE cores).
However, we use considerably fewer cores for MatMul kernels (312 \textit{vs.} 384), while
the remaining (390-312=78) AIE cores are used to run Add kernels.
Also notice that CHARM has only MatMul kernels.
Therefore, our solution is also able to achieve less AIE core power consumption (25.62 W \textit{vs.} 26.95 W), because the cores that run the fp32 Add kernels remain idle most of the time (Table \ref{tb:single_AI_engine_results}).
However, our implementation uses more memory banks than CHARM (3138 \textit{vs.} 3086 out of 3200 available), which leads to higher data memory power consumption (18.21 W \textit{vs.} 16.74 W).
When computing the total AIE power as the summation of AIE core power and data memory power \cite{AIE_XPE}, we observe that our 13$\times$4$\times$6 design exhibits slightly higher power than CHARM (43.83 W \textit{vs.} 43.69 W).
Hence, our highest throughput solution presents also \textbf{20.4\%} higher energy efficiency compared to CHARM.
We note here that our method of input broadcasting and output adder tree reduction, utilizes efficiently the available PLIOs (79\% for 13$\times$4$\times$6).
On the contrary, CHARM severely under-utilizes the device's PLIOs (only 41\%), which acts as a performance bottleneck for their design.
Finally, we observe a very small DMA usage of 18 banks due to the \emph{``T"--like} shapes of pattern $P1$ (see Fig. \ref{fig:placement_patterns}), as expected.

\subsubsection{MaxEVA \textit{vs.} state-of-the-art CHARM for int8 precision}
Since int8 CHARM implementation is not open-sourced, we perform a qualitative comparison of performance.
In \cite{dac23automm} the authors report MatMul throughput of 28.15 TOPs for int8 CHARM design, when operating at 1 GHz frequency.
To fairly compare with our results, we scale the aforementioned value to 1.25 GHz (our frequency), thus becoming  35.19 TOPs.
In contrast, MaxEVA presents int8 maximum throughput of 77.01 TOPs, which is \textbf{2.19$\times$} higher than CHARM (Table \ref{tb:MaxEVA_results_int8}).
To get more confidence, we do a similar qualitative comparison for fp32 results. When scaling for fp32, we get a CHARM performance of 4342.33 GFLOPs at 1.25GHz.
But our experimental results in Table \ref{tb:MaxEVA_results_fp32} show a performance of 4504.46 GFLOPs for fp32 CHARM implementation.
This small performance difference of 3.73\% is expected because the authors in \cite{dac23automm} measure the end-to-end performance on the VCK190.
Thus, they experience sources of performance degradation, including the required zero padding \cite{dac23automm}.
However, this small difference indicates that our experiments are accurate, and also validates our $\sim$2.19$\times$ performance gain for int8 over CHARM.
This substantial performance gain is because CHARM utilizes only 192 AIE cores (48\%) for int8, due to routing congestion issues \cite{dac23automm}.
On the contrary, MaxEVA utilizes efficiently the entire AIE array, by mapping 390 cores (97.5\%) for the highest throughput design (row 1 in Table \ref{tb:MaxEVA_results_int8}).
Finally, we note that due to the absence of open-source code, power for int8 CHARM cannot be calculated through the XPE tool, thus we are not able to present energy efficiency comparison.

\subsubsection{Placement Patterns Comparison}
To provide a comprehensive evaluation of the proposed placement patterns, we show the two top-ranked solutions for each pattern in Tables \ref{tb:MaxEVA_results_fp32}, \ref{tb:MaxEVA_results_int8} (rows 1--4).
Based on the results, in general, we observe that pattern $P2$ has higher total AIE core and memory usage compared to $P1$, because it uses more Add kernels.
However, higher AIE core usage does not necessarily lead to higher core power consumption.
For instance, pattern $P2$ 10$\times$3$\times$10  design utilizes the entire AIE array (400 cores, feasible routing for pattern $P2$ since no DMA is used), but exhibits lower AIE core power than $P1$  13$\times$4$\times$6 design (25.54 W \textit{vs.} 25.62 W and 47.44 W \textit{vs.} 48.65 W for fp32 and int8, respectively).
This is attributed to the fact that $P2$ has fewer MatMul kernels than $P1$ (300 \textit{vs.} 312), and more cores that run Add kernels which remain mostly idle (100 \textit{vs.} 78).
However, when also including the memory power, the total power consumption depends on the number of memory banks used as well.
In particular, when comparing 10$\times$3$\times$10 with 13$\times$4$\times$6, the former shows slightly higher total power for fp32 (44.66 W \textit{vs.} 43.83 W), while for int8 its power is lower (65.52 W \textit{vs.} 66.83 W) than the latter.

In general, we observe from Tables \ref{tb:MaxEVA_results_fp32}, \ref{tb:MaxEVA_results_int8} that the higher the number of MatMul kernels, the higher the throughput.
However, this does not always hold true.
For instance, for int8 precision, the 10$\times$3$\times$10 design presents slightly higher throughput than 11$\times$4$\times$7 (76.08 \textit{vs.} 75.67 TOPs), despite the fact that it has fewer MatMul kernels (300 \textit{vs.} 308).
This very small performance difference ($<$1\%) is due to memory conflicts (leading to a few stalls), caused by
dissimilarities in buffer optimizations from the AIE PnR tool \cite{AI_Engine_programming_environment}.


To quantify the effect of DMA usage on MatMul performance, we also implement the highest \textit{common} solution (same number of MatMul kernels) between our two placement patterns (Tables \ref{tb:MaxEVA_results_fp32}, \ref{tb:MaxEVA_results_int8}, rows 5--6).
In particular, when comparing 12$\times$4$\times$6 ($P1$) with 12$\times$3$\times$8 ($P2$), which both have 288 MatMul kernels, we notice that throughput is higher in $P2$ for both precisions.
For instance, for int8, $P1$ attains 71.25 TOPs, while $P2$ achieves 72.93 TOPs.
This is attributed to the DMA resources used in $P1$, which increase latency compared to $P2$, where no DMA is used.
However, from Tables \ref{tb:MaxEVA_results_fp32}, \ref{tb:MaxEVA_results_int8} we observe that $P2$ has higher energy efficiency for int8 (1.153 \textit{vs.} 1.147 TOPs/W), while for fp32 the opposite occurs (123.58 \textit{vs.} 123.68 GFLOPs/W).
This arises from the fact that cores running Add kernels remain idle for significantly fewer cycles for int8 compared to fp32 (Table \ref{tb:single_AI_engine_results}).
To this end, we observe a higher percentage difference of total power for fp32 when comparing the aforementioned $P1$ and $P2$ solutions (40.68 W \textit{vs.} 42.28 W for fp32, and 
62.13 W \textit{vs.} 63.24 W for int8).
We notice that although in most cases $P2$ and $P1$ present higher energy efficiency for int8 and fp32, respectively, this relationship is complicated and depends on the number of MatMul kernels, the total cores used, as well as the memory banks and switch routing (as optimized by the AIE PnR tool).

Overall, throughout all design points, 13$\times$4$\times$6 ($P1$) exhibits both highest throughput (5442.11 GFLOPs, \textbf{20.8\%} over CHARM) and energy efficiency (124.16 GFLOPs/W, \textbf{20.4\%} higher than CHARM), for fp32 precision.
However, for int8, 13$\times$4$\times$6 ($P1$) has the highest throughput (77.01 TOPs, \textbf{2.19$\times$} higher than CHARM), while 10$\times$3$\times$10 ($P2$) exhibits the greatest energy efficiency (1.161 TOPs/W).
Finally, all of our optimized designs present very high resource utilization, using up to 100\% AIE cores, 99.8\% AIE memory and 82.1\% PLIOs.




\begin{figure}[t]
 \vspace{-0.60cm}
\centering
\subfloat[fp32]{\includegraphics[width=0.50\linewidth]{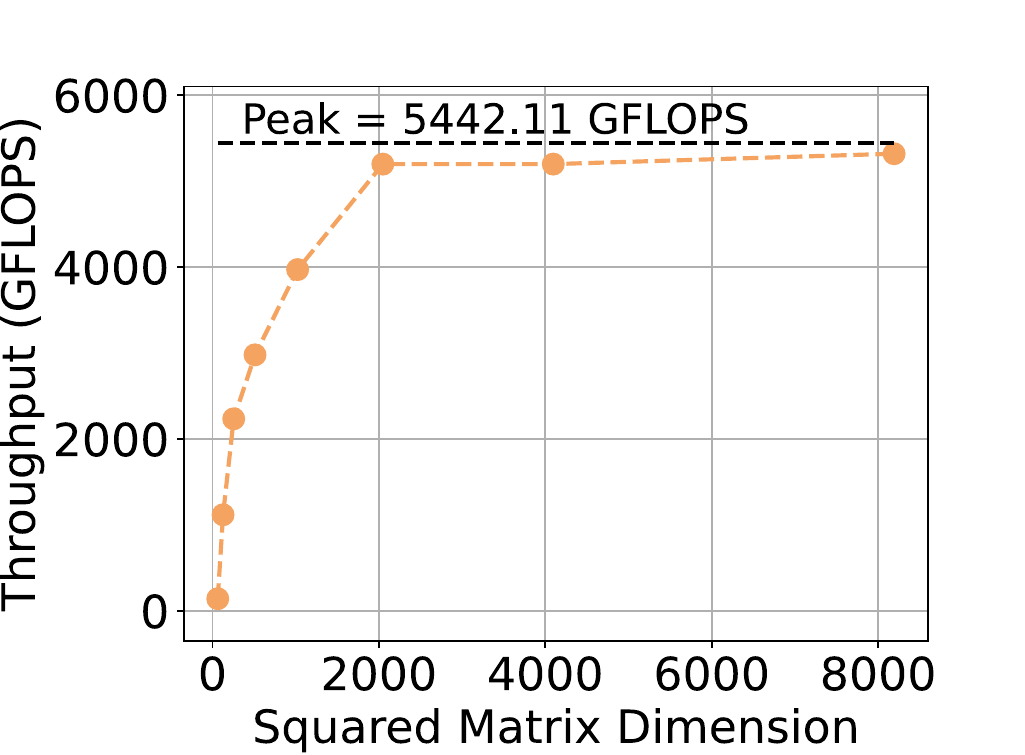}
\label{fig:throughput_mat_size_fp32}}
\subfloat[int8]{\includegraphics[width=0.48\linewidth]{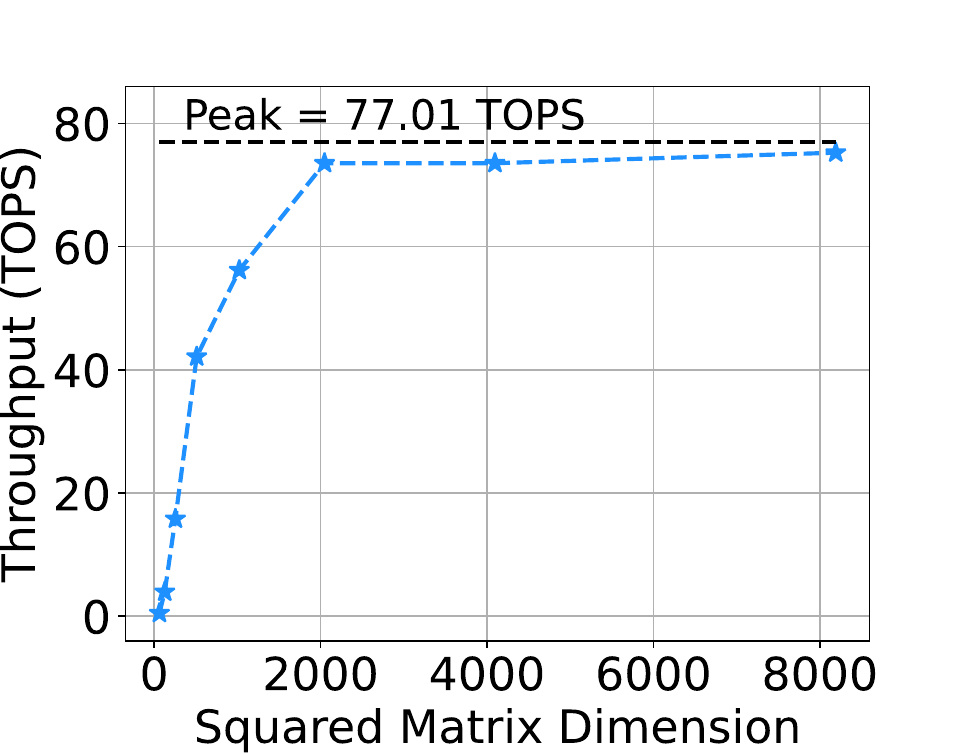}
\label{fig:throughput_mat_size_int8}}
\caption{Variation of throughput for different square matrix sizes for the 13$\times$4$\times$6 design.}
\label{fig:throughput_per_matrix_size}
\vspace{-0.35cm}
\end{figure}

\subsubsection{Variation of Performance under Different Matrix Sizes}
We also explore the performance variation when altering the input matrix sizes (as powers of two) of the highest throughput design (Fig. \ref{fig:throughput_per_matrix_size}).
The throughput is estimated by supposing that tiling is performed in PL for large matrix sizes, and also the PL does not cause any stalls (commonly attained in practice \cite{dac23automm}).
As expected, we observe that as the matrix size increases, the throughput also increases, and for large enough matrices it converges to its peak value.
This is ascribed to zero padding in matrices such that they fit the native MatMul size of the 13$\times$4$\times$6 design.
In particular, the 13$\times$4$\times$6 design is able to perform a MatMul of 416$\times$128$\times$192 and 416$\times$512$\times$192 size for fp32 and int8, respectively.
To this end, we notice that for square matrices larger than $\sim$2K$\times$2K$\times$2K, less padding is needed throughout tiling, leading to almost peak performance.

Going a step further, we estimate the performance of full DNN inference, under the same assumptions as above.
More specifically, when considering the MLP used in \cite{charm2023fpga}, MaxEVA achieves a throughput of 4735.94 GFLOPs.
In contrast, when scaling the reported results from \cite{charm2023fpga} to 1.25 GHz, we get  3670.88 GFLOPs, showcasing a higher MaxEVA performance of \textbf{29\%} over CHARM. 
Finally, we note that our work can be extended in straightforward fashion to other special cases of MatMul, e.g., Matrix-Vector, which we leave as future work.








\section{Conclusion}
\label{sec:Conclusion}

The Versal AIE architecture introduces a new paradigm in reconfigurable computing, while posing several 
unique design challenges.
To resolve these new challenges, we propose the novel MaxEVA framework.
MaxEVA successfully maximizes the efficiency of MatMul on Versal AIE, by effectively leveraging the AIE characteristics and addressing performance bottlenecks from various perspectives.
Our experimental results show remarkable performance gains over the state-of-the-art design of up to \textbf{2.19$\times$} higher throughput and \textbf{20.4\%} greater energy efficiency.
The MaxEVA framework is generalizable to any Versal AIE platform and MatMul-based DL workloads.







\section*{Acknowledgment}
This work was supported in part by the 
National Science Foundation
CCF Grant No. 2107085, iMAGiNE -
the Intelligent Machine Engineering Consortium at UT Austin, and a UT
Cockrell School of Engineering Doctoral Fellowship.

\clearpage

\bibliographystyle{ieeetr}
\small{\bibliography{bibtex}}

\end{document}